\documentclass[%
 preprint,
 amsmath,amssymb,
 aps,nofootinbib
]{revtex4-2}

\usepackage{graphicx}
\usepackage{dcolumn}
\usepackage{bm}
\usepackage{booktabs}
\usepackage{hyperref}

\newcommand{\VS}{\mathrm{VS}}

\newcommand{\Mss}{M_{\rm spin\text{-}0}}
\newcommand{\gt}{\tilde g}

\newcommand{\Lp}{\mathcal{L}_+}

\begin{document}

\preprint{}

\title{Analytic Boundaries of Infinite-Spin-Tower Amplitudes from Hidden Zero}

\author{Long-Qi Shao}
\email{longqi.shao@df.unipi.it}
\affiliation{Department of Physics, University of Pisa and INFN,
Largo Pontecorvo 3, I-56127 Pisa, Italy}
\author{Alessandro Vichi}
\email{alessandro.vichi@unipi.it}
\affiliation{Department of Physics, University of Pisa and INFN,
Largo Pontecorvo 3, I-56127 Pisa, Italy}

\date{\today}







\vspace{1.5em}

\begin{center}
\begin{minipage}{\textwidth}

\begin{center}
\maketitle
\textbf{Abstract}
\end{center}

{\small \hspace{1.5em}
We study the general form of meromorphic amplitudes that are compatible with unitarity, analyticity, crossing symmetry, polynomial boundedness and the hidden-zero and corresponding splitting conditions. These amplitudes are infinite-spin-tower (IST) amplitudes and are characterized by the distribution of poles. With infinitely many equidistant poles, the IST amplitudes reduce to Veneziano amplitudes. We construct such unitary amplitudes in a primal way (rule in) and find the bounds analytically. The allowed region we derive is smaller than, but close to, the region allowed by the positivity bounds (rule out). We argue that, with the conditions imposed, the analytic boundary we derive is the largest possible boundary in the primal construction of meromorphic amplitudes. 
This type of IST amplitude can also be extended to the fully crossing-symmetric case related to the Virasoro-Shapiro amplitude. We found from the IST amplitudes that graviton pole imposes unitarity constraints on UV spectrum.}
\end{minipage}
\end{center}

\vspace{1.5em}
\noindent\rule{6.5cm}{0.4pt}\\
$\,^\star$
\href{mailto:longqi.shao@df.unipi.it}{longqi.shao@df.unipi.it}\\
$\,^\dagger$ \href{mailto:alessandro.vichi@unipi.it}{alessandro.vichi@unipi.it}\\

\newpage

\tableofcontents

\section{Introduction}\label{sec:intro}

Bootstrap methods have proved powerful in constraining different theories~\cite{Adams:2006sv,Rattazzi:2008pe,El-Showk:2012cjh,Poland:2018epd,Kos:2016ysd,Simmons-Duffin:2015qma,Arkani-Hamed:2020blm,Guerrieri:2021ivu,Guo:2025fii,Arkani-Hamed:2026rsz}; identifying the theories that lie at kinks and on the boundaries of the allowed regions is therefore usually important. The bootstrap assumes only the most basic principles, including unitarity, causality, analyticity, and crossing symmetry. It is also interesting to include more specific properties in different contexts and investigate how uniquely a theory is determined by these \textit{ad hoc} assumptions. One particularly interesting context is the question of the uniqueness of the Veneziano amplitudes~\cite{Veneziano:1968yb,Baker:1970vxk,Caron-Huot:2016icg,Cheung:2022mkw,Geiser:2022exp,Cheung:2023uwn,Chiang:2023quf,Cheung:2024obl,Arkani-Hamed:2023jwn}, namely, which conditions are sufficient to determine string amplitudes uniquely.

In the study of $\mathrm{Tr}\,\phi^3$ amplitudes~\cite{Arkani-Hamed:2023swr,Arkani-Hamed:2024nhp}, it was found that higher-point amplitudes exhibit ``hidden zeros'' on a special kinematic locus. Along with these ``hidden zeros,'' the amplitudes also factorize into lower-point amplitudes when this kinematic locus is approached, with one non-planar variable on the hidden-zero locus taken to be nonzero. Owing to the close connection between $\mathrm{Tr}\,\phi^3$ theory and other colored theories, these properties have been studied extensively~\cite{Berman:2025owb,Wan:2026pjq,Arkani-Hamed:2024jbp,Bartsch:2024amu,Cheung:2024uhn,Li:2024qfp,Rodina:2024yfc,Feng:2025ofq,De:2025bmf,CarrilloGonzalez:2026lnu}.

The hidden zeros and splitting conditions combined with basic principles were studied to bootstrap the EFT~\cite{Berman:2025owb} whose leading order interaction is $\mathrm{Tr}\,\phi^3$ theory. It was found that, once one requires only finitely many spins at the leading energy level, the allowed region shrinks to a small island, leaving room only for the Veneziano amplitude. The numerical bootstrap results suggest that these properties are sufficient to determine the Veneziano amplitude uniquely. This uniqueness resulting from the hidden-zero condition was later demonstrated analytically in the presence of infinitely many energy levels~\cite{Wan:2026pjq}.

In this paper, we use a primal approach to analytically construct meromorphic $s,t$ crossing-symmetric amplitudes that satisfy the aforementioned bootstrap principles and hidden-zero conditions. This is a rule in method. The boundary we derive lies close to the positivity bounds. We prove that the analytic boundary we find is the optimal boundary for meromorphic amplitudes. The remaining gap between the primal bound and the positivity bounds is likely to be caused by slow numerical convergence with respect to number of nonlinear constraints coming from splitting conditions. Namely, it is difficult to implement fully these nonlinear constraints in positivity bounds, similar to the problem of fully implementing crossing symmetry.

The amplitudes we construct are called infinite-spin-tower (IST) amplitudes~\cite{Caron-Huot:2020cmc,Huang:2022mdb,Berman:2025owb}, since they contain states of arbitrarily high spin at each energy level. The IST amplitudes can be generalized to the fully $s,t,u$ crossing-symmetric case. We note that the same IST amplitudes were recently found in Ref.~\cite{Berman:2026ezk}. These amplitudes were previously regarded as toy amplitudes because states of arbitrarily high spin at a single mass level were thought to conflict with locality. However, these IST amplitudes are dominated by low-spin states~\cite{Bern:2021ppb}. The coupling constants of the high-spin particles are exponentially small. Whether the existence of these high-spin particles with negligible coupling constants is in tension with nature remains unknown~\cite{Weinberg:1965nx,Bousso:1999xy,Biswas:2012bp,Maldacena:2022ckr}. A worldsheet description of IST amplitudes was discussed in Ref.~\cite{Wang:2024jhc}, suggesting that IST amplitudes need not necessarily be unphysical. Recently, a construction of IST amplitudes was shown to be free of divergences to all loop orders~\cite{Calisto:2026cdy}. Nevertheless, it remains interesting to study these amplitudes because they may lie marginally within the positivity bounds, namely, on the boundaries of the allowed region.

Finally, The exponentially decreasing couplings are reminiscent of the extremal spectra encountered in the S-matrix bootstrap~\cite{Albert:2023seb,Berman:2024eid,Albert:2024yap} under the assumption of meromorphic amplitudes. The extremal boundary is reached by maximizing the coupling of the lowest-lying spin-two state. The spectrum on the extremal boundary tends to contain one nonlinear leading Regge trajectory together with many spurious high-spin states. These spurious states also have exponentially small coupling constants. Although it remains an open question whether closed-form amplitudes lying on these boundaries can be found, the simplest IST amplitudes usually lie at the kink connected to the extremal boundary. This suggests that IST amplitudes, or deformations of IST amplitudes, may provide candidate extremal solutions.

In Sec.~2, we introduce IST amplitudes as solutions to the functional equalities required by the splitting condition. In Sec.~3, we determine the analytic boundaries of the IST amplitudes within the positivity bounds. In Sec.~4, we show analytically the IST amplitudes constructed are unitary. In Sec.~5, we generalize the IST construction to the fully $s,t,u$ crossing-symmetric case.

\section{Hidden zeros and splitting condition}
The hidden-zero condition refers~\footnote{We will write down the hidden-zero and splitting conditions directly, and refer the interested reader to~\cite{Arkani-Hamed:2023swr,Arkani-Hamed:2024nhp} for their derivation.} to zeros of color-ordered amplitudes of $\mathrm{Tr}\,\phi^3$ theory in specific kinematic slices~\cite{Arkani-Hamed:2023swr,Arkani-Hamed:2024nhp}. When one of the kinematic variables deviates from the slice, the higher-point amplitude factorizes into a product of lower-point amplitudes. It turns out these are “stringy” conditions that uniquely determine certain types of amplitudes~\cite{Berman:2025owb,Wan:2026pjq}---Veneziano amplitudes and IST amplitudes.

We start with general EFT ansatz for color ordered 4-point amplitudes whose leading order amplitude is $\mathrm{Tr}\,\phi^3$ amplitude
\begin{equation}
    A_4(s,t)
= (s+t)\left( -\frac{g^2}{st} + \sum_{k=0}^{\infty} \sum_{0 \leq q \leq k} a_{k,q}\, s^{k-q}\, t^{q} \right).
\end{equation}
The first term is the leading-order massless scalar exchange term and the second term contains contact interactions. The factor $(s+t)$ is required by the hidden-zero condition for the 4-point amplitude, which means that when $s+t=0$, the amplitude vanishes. Hereafter, we will set $g=1$ for convenience. The hidden-zero condition for the 5-point amplitude does not matter here, but the corresponding splitting condition gives a nontrivial relation between the 4-point amplitude and the 5-point amplitude.
\begin{equation}
    A_5\big|_{X_{14}=X_{13}+X_{24}} =
A_4(X_{13},X_{25})A_4(X_{24},X_{35}),  \label{fac}
\end{equation}
where $X_{ij}=(p_i+p_{i+1}+\ldots+p_{j-1})^2$ are the kinematic variables and $p_i$ is an external momentum. Take a cyclically rotation
\begin{equation}
    A_5\big|_{X_{25}=X_{24}+X_{35}} =
A_4(X_{24},X_{31})A_4(X_{35},X_{41}),  \label{splitting}
\end{equation}
Since $A_5$ is cyclic invariant, in the kinematic slice $X_{14}=X_{13}+X_{24}$ and $X_{25}=X_{24}+X_{35}$, we can subtract $A_5$ and get
\begin{equation}
A_4(X_{13},X_{24}+X_{35})A_4(X_{24},X_{35})=A_4(X_{24},X_{31})A_4(X_{35},X_{13}+X_{24}).
\end{equation}
Making the substitutions $X_{13}=a, X_{24}=b,X_{35}=c$ gives
\begin{equation}
A_4(a,b+c)A_4(b,c)=A_4(a,b)A_4(a+b,c).  \label{associate}
\end{equation}
For meromorphic amplitudes, this can always be solved by the following form, see appendix~\ref{appA}
\begin{equation}
\boxed{\;A_4(s,t)\;=\;C\,\frac{f(s)\,f(t)}{f(s+t)},\;}
\label{eq:gaussbeta}
\end{equation}
where $C$ is a constant. The class of solution is closed under products of $f$. 

Suppose the amplitudes have simple poles at $s=\{0,\mu_1,\mu_2,\ldots,\mu_n\}$, further impose $s,t$ crossing symmetry, causality, planarity(no u-channel pole), polynomial boundedness in complex $s$ plane and Eq.~\eqref{eq:gaussbeta}, the only meromorphic amplitude we can write down is the IST amplitudes~\footnote{There is still an entire function ambiguity for solutions satisfying Eq.~\eqref{associate}, imposing polynomial boundedness will eliminate the ambiguity. In the study of deformation of open string amplitudes, one starts from ansatz $\frac{a_n\,s t+b_n(s+t)+c_n}{(\mu_n-s)(\mu_n-t)}$. But given the denominator, Eq.~\eqref{eq:gaussbeta} fixes the numerator to be $(\mu_N-s-t)$ up to a constant.}
\begin{equation}
    -\frac{s+t}{st}\prod_{N=1}^n\frac{C_N(\mu_N-s-t)}{(\mu_N-s)(\mu_N-t)}.  \label{GIST}
\end{equation}
It has infinitely many states in each massive energy level because of the $t$ dependence in the denominator. It is bounded by $s^1$ in the Regge limit, compatible with the once-subtracted dispersion relation used in numeric bootstrap~\cite{Berman:2025owb}. The pole distribution $\{\mu_N\}$ characterizes the different IST amplitudes. Value of $C_N$ will not matter in most cases. 

So far we have not required unitarity for Eq.~\eqref{GIST}. The unitarity in this paper refers to partial wave unitarity for meromorphic amplitudes, i.e. at each pole $s=k$, the residues can be Legendre expanded in $\cos\theta=1+2t/k$ with positive coefficients. We will show below how unitarity affects the distribution of poles. 

A special choice of Eq.~\eqref{GIST} is $C_N=N$, $\mu_N=N$, ($\mu_1$ was set to be unit of mass)
\begin{equation}
    A(s,t)= -\frac{s+t}{st}\prod_{N=1}^n\frac{N(N-s-t)}{(N-s)(N-t)}.  \label{FP}
\end{equation}
From now on, we will only talk about four-point colour-ordered meromorphic amplitudes; therefore, we use $A$ instead of $A_4$.  The amplitudes Eq.~\eqref{FP} has equidistant poles, it  will be proven to satisfy unitarity below. But there are more IST amplitudes that are unitary without equidistant poles. For $n\to\infty$, Eq.~\eqref{FP} is just the infinite product representation of the Veneziano amplitude. All $t$ dependence in the denominator will be canceled, leaving polynomials of $t$ when calculating residues. $\prod_N C_N$ does not matter when there are finitely many terms. When there are infinitely many terms, $C_N$ has to be $N$ to make the infinite product converge and non-vanishing.
\begin{figure}[htp]
    \centering
    \includegraphics[width=0.7\linewidth]{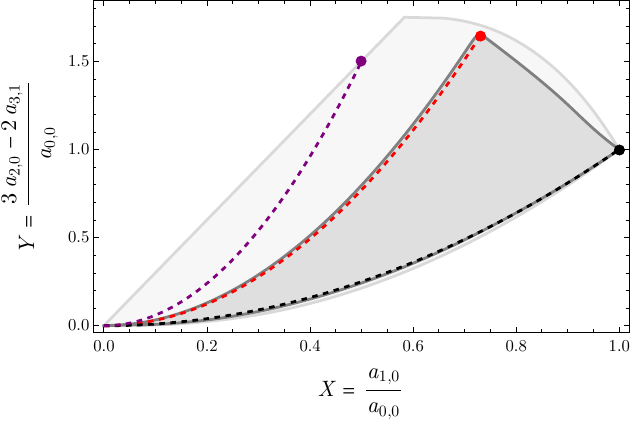}
    \caption{The allowed region from positivity bounds~\cite{Berman:2025owb}. The light grey area is the allowed region before implementing nonlinear splitting constraints Eq.~\eqref{fac}, the dark grey region is the region after implementing splitting constraints. The red dashed line is Veneziano amplitude with Regge slope $\alpha'\in[0,1]$, the black dashed line is IST amplitude Eq.~\eqref{FP} with $n=1$ and $\mu_1\in[1,\infty]$. The purple dashed line is massive scalar vector, which is ruled out by the splitting condition.}
    \label{BEF}
\end{figure}
\section{Analytic boundary}
In Fig.~\ref{BEF}, the positivity bounds (rule out) on 4-point color ordered amplitudes whose leading order contribution is given by the $\mathrm{Tr}\,\phi^3$ amplitude were derived~\cite{Berman:2025owb}. After implementing the nonlinear constraints from the splitting condition Eq.~\eqref{splitting}, the allowed region shrinks from the light grey area to the dark grey area. The red dot is the Veneziano amplitude with $\alpha'=1$,
\begin{equation}
    A_V(s,t)=\frac{\Gamma(-\alpha's)\Gamma(-\alpha't)}{\Gamma(-\alpha's-\alpha't)}.
\end{equation}
As the Regge slope $\alpha'$ varies from 1 to 0, it traces out the red dashed line. The black dashed line is IST amplitude with one massive pole
\begin{equation}
    A_{IST}=-\frac{s+t}{st}\frac{(1-s-t)}{(1-s)(1-t)}.
\end{equation}
With the mass scale from 1 to infinity, it traces out the black dashed line. After implementing the nonlinear constraints, the standard convex hull argument no longer holds. One cannot add two unitary amplitudes with a positive weight anymore. Because it breaks the nonlinear functional relation Eq.~\eqref{associate}. Given one more spectrum input that there are finitely many spin states in the lowest-lying pole, the allowed region immediately shrinks to the red dot --- the Veneziano amplitude. What are the theories in the dark grey region, and why is the dark grey region simply connected given that additivity is lost? We will answer these questions by constructing analytic amplitudes that live inside the dark grey region.

These numerical results already suggest that all other points in the dark grey region have infinite spin in the first massive pole. In order to compare with the numerical positivity bounds in Fig.~\ref{BEF}, we define the coordinates
\begin{equation}
    X=\frac{a_{1,0}}{a_{0,0}}, \quad Y=\frac{3a_{2,0}-2a_{2,1}}{a_{0,0}}. \label{eq:PC}
\end{equation}
For the most general IST amplitudes Eq.~\eqref{GIST}, their $X,Y$ coordinates are surprisingly simple
\begin{equation}
    X=\frac{\sum_N\mu_N^{-3}}{\sum_N\mu_N^{-2}},\quad Y=\sum_N \mu_N^{-2}.  \label{coord}
\end{equation}
where $\mu_N$ are the locations of the poles. As mentioned above, the distribution of poles is the only parameter we can tune.

Now we need to understand the unitarity constraints on the distribution of the poles. In general, it is a complicated and open problem to know how unitarity dictates the spectrum. We found only a necessary condition for unitarity in this IST setup:

\textit{The separation between adjacent energy levels must be at least as large as the first energy level, i.e. the mass gap.}

\begin{figure}[htp]
    \centering
    \includegraphics[width=1\linewidth]{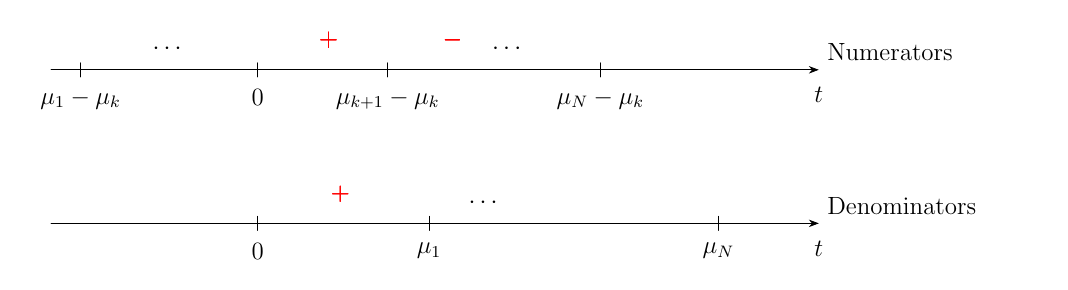}
    \caption{Energy separation of the IST families. Energy separation be larger than $\mu_1$ is necessary for unitarity.}
    \label{fig:energy_separation}
\end{figure}

The argument for this claim is as follows: the unitarity for a meromorphic amplitude requires that the residue at every pole can be expanded in terms of Legendre polynomials with positive coefficients
\begin{equation}
    \text{Res}_{s=k}A(s,t)=\sum_j c_{k,j}P_j(1+2t/k).
    \label{Legendre}
\end{equation}
Since all $c_{k,j}$ are positive and $P_j(1+2t/k)$ is positive when its argument is larger than 1, i.e. when $t>0$, the residue should be positive when $t>0$. This is a necessary condition for unitarity. Now the residues for IST amplitudes are
\begin{equation}
    \text{Res}_{s=\mu_k}A(s,t)\propto \prod_N\frac{\mu_N-\mu_k-t}{\mu_N-t}.
\end{equation}
The zeros of the denominators are $\{0,\mu_1,\mu_2,\mu_3\ldots,\mu_N\}$. The zeros of the numerators are $\{\mu_1-\mu_k,\ldots,\mu_{k-1}-\mu_k,0, \mu_{k+1}-\mu_k,\ldots,\mu_N-\mu_k\}$, i.e. shifted by a constant $\mu_k$. This is shown in Fig.~\ref{fig:energy_separation}. Starting at $t=0$ and increase $t$, the sign of residue changes whenever it hits a zero (numerator or denominator). If $\mu_{k+1}-\mu_k<\mu_1$, when $t$ reaches $\mu_{k+1}-\mu_k$, the residue changes sign and breaks unitarity. On the contrary, if $\mu_{k+1}-\mu_k\geq\mu_1$, it does not break unitarity. Above $\mu_1$, the Legendre expansion Eq.~\eqref{Legendre} diverges, so it does not matter what happens for $t>\mu_1$.

Based on this, we can distribute the poles of IST amplitudes in the following way. First, we put all the massive poles at infinity; they are irrelevant in the IR. Indeed, a pole with infinite mass contributes only a constant factor to the product Eq.~\eqref{GIST}. Then we move one pole to the IR; this pole gives the mass gap. We can always set it to be 1 (mass units). Then we move the second pole to the IR; the lowest energy it can reach is $\mu_2=2$ because of unitarity requirement stated above. Then we move the third pole to $\mu_3=3$. Continuing this procedure traces out a line,
\begin{equation}
    X=\frac{H_n^{(3)}+(Y-H_n^{(2)})^{3/2}}{Y} \quad \text{for} \quad H_n^{(2)}\leq Y \leq H_{n+1}^{(2)}\,, \label{curve}
\end{equation}
where $H_n^{(m)}=\sum_{k=1}^{n} k^{-m}$ is the harmonic number. This corresponds to the most economical choice with the fewest poles located at $\{0,1,2,\ldots,n, r\}$ with $r\geq n+1$. The amplitudes are 
\begin{equation}
    A(s,t)=-\frac{s+t}{st}\prod_{N=1}^n\frac{N(N-s-t)}{(N-s)(N-t)}\frac{r(r-s-t)}{(r-s)(r-t)}. \label{bamp}
\end{equation}
We can show that all other distributions of poles will lead to a smaller $X$ for fixed $Y$, see Appendix~\ref{appB}. The unitarity of amplitudes constructed this way will be shown in the next section. 

\begin{figure}
    \centering
    \includegraphics[width=0.9\linewidth]{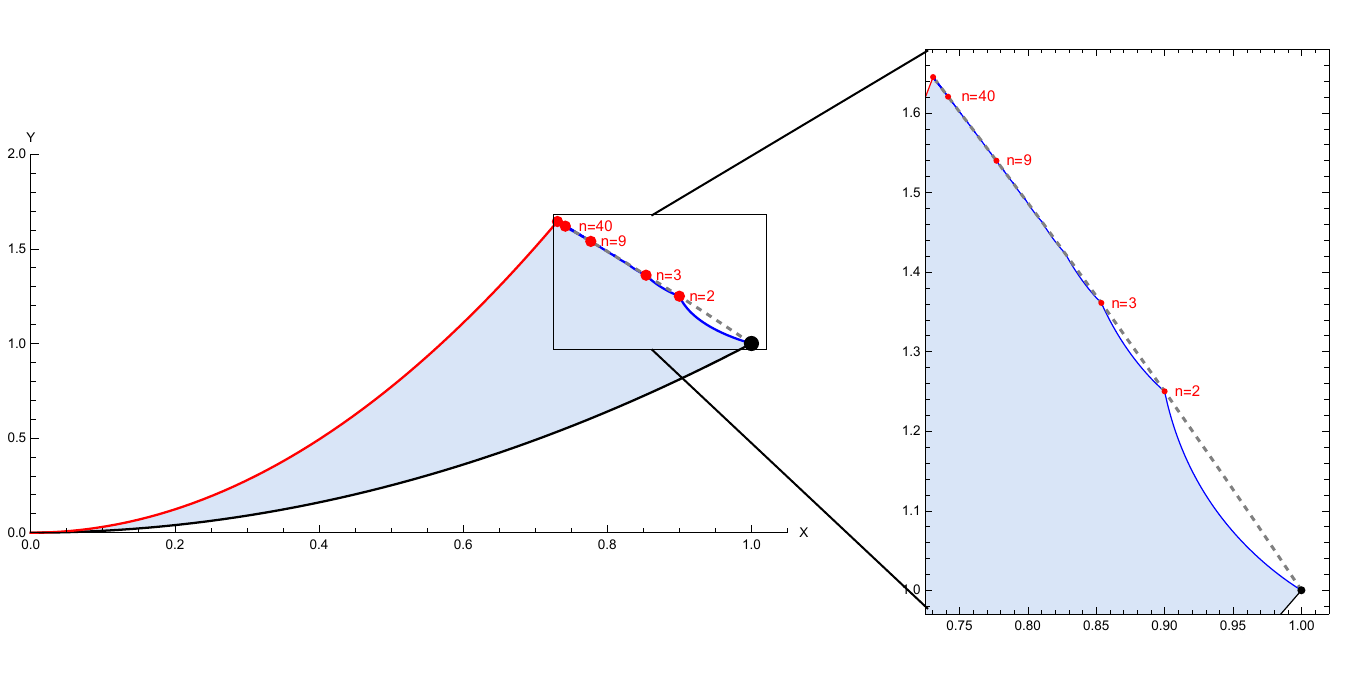}
    \caption{The blue region constructed by IST amplitudes. The red dots denote IST amplitudes with poles distributed at $\{1,2,\ldots,n\}$. For adjacent dots, there is a continuous curve connecting the dots, they correspond to IST amplitudes with poles distributed at $\{1,2,\ldots,n,r\}$ with $r\geq n+1$.}
    \label{PC2}
\end{figure}

The continuous curve Eq.~\eqref{curve} is shown in Fig.~\ref{PC2}.
With these curved lines and their mass-scaling curves (the curves derived by multiplying each pole by a constant larger than 1), we can see that the allowed region is simply connected now.

There is still a gap between the blue line and the dashed line denoting SDPB result~\cite{Simmons-Duffin:2015qma,Berman:2025owb} in Fig.~\ref{PC2}. As shown in Appendix~\ref{appB}, the blue line gives the largest $X$ for fixed $Y$, so it is the best one can find with IST amplitude constructions mentioned above.
Notice that if we take the pole distribution to be $\{0,1,2,\ldots,n\}$, then $X=H_n^{(3)}/H_n^{(2)}$ and $Y=H_n^{(2)}$. If we just do analytic continue in $n$ such that it can be non-integer, the curve will trace out a larger boundary that is much closer to the rule-out boundary from SDPB, but the corresponding amplitudes~\footnote{One can turn the amplitude from the product form into a Gamma function.} are problematic. They have $u$ channel pole which violates planarity of the color ordered amplitudes. They have poles in physical region $|\cos\theta|=|1+2t/s|\leq1$. They are ruled out in our primal bootstrap setup.

There are two possibilities for how this gap arises. The first is that there may be an accumulation point of the poles. This scenario is different from our constructed IST amplitudes. It is possible such infinite product amplitudes unitary, like Coon amplitudes~\cite{Figueroa:2022onw,Chakravarty:2022vrp,Bhardwaj:2022lbz,Wang:2024wcc,Ambrosino:2026ddt}~\footnote{The Coon amplitude is not directly relevant here because it does not satisfy Eq.~\eqref{eq:gaussbeta}.}. Although such amplitudes look meromorphic, they exhibit a branch point at the accumulation point.

The other possibility is that the positivity bounds derived from SDPB is not optimal, it is not possible to implement fully the nonlinear relation Eq.~\eqref{fac} numerically. This is the same with the case of $s,t$ crossing symmetry. When doing positivity bounds, one needs to implement null constraints~\cite{Tolley:2020gtv,Caron-Huot:2020cmc} derived from crossing symmetry, but there are infinite many of them. In the numerics, one should impose increasingly many null constraints until the bounds converge. Regarding the nonlinear relation from Eq.~\eqref{fac} here, it has not been shown that the rule-out boundary has converged~\cite{Berman:2025owb}. After implementing more nonlinear constraints from Eq.~\eqref{associate}, we observe that rule-out boundary move closer to our analytic boundary. The gap is still there because the convergence is slow with respect to number of nonlinear relations. And it becomes increasingly more difficult to impose more nonlinear constraints due to technical issues discussed in~\cite{Berman:2025owb}. Therefore, if one can improve the exclusion plot, we may be able to close the gap.

\section{Unitarity for the IST amplitudes}\label{subsec:positivity}

The unitarity for the IST amplitudes depends on the distribution of the poles. We will first give an analytic proof of unitarity for IST amplitudes with equidistant poles. Then we will show analytic proof for the IST amplitudes living on the boundary with poles on $\{0,1,2,\ldots,n,r\}$ where $r>n+1$~\footnote{We thank Ragan Ewing for providing an analytic proof of unitarity on the pole $s=r$, obtained with the assistance of AI model GPT-5.6 Sol.}. We will define the set of meromorphic functions that has the unitarity property
$\Lp:=\Bigl\{\,f=\sum_{j}c_j\,P_j(\cos\theta),\ \ \cos\theta\in[-1,1]\ :\ c_j\ge0,\ \ \sum_{j}c_j<\infty\,\Bigr\}.$

\subsection{IST amplitudes with equidistant poles}
Consider the amplitude with equidistant poles:
\begin{equation}\label{eq:amplitude}
  A(s,t) = -\frac{s+t}{s\,t}\prod_{N=1}^{n}\frac{N(N-s-t)}{(N-s)(N-t)}\,,
\end{equation}
which has simple poles at $s=\{0,1,\dots,n\}$. A direct residue computation at $s=k$ gives
\begin{equation}
  R_{n,k}(t) = \underbrace{\binom{t+k}{k}}_{\displaystyle V_k(t)}
  \;\cdot\;\underbrace{\prod_{l=0}^{k-1}\frac{n-l}{n-l-t}}_{\displaystyle C_{n,k}(t)},  \label{PF}
\end{equation}
where $\binom{t+k}{k}=\frac{(t+1)(t+2)\cdots(t+k)}{k!}$. The function $V_k = \binom{t+k}{k}$ is a polynomial of degree $k$ in $t$. It is the residue of the Veneziano amplitude 
$\Gamma(-s)\Gamma(-t)/\Gamma(-s-t)$ at $s=k$, 
whose partial wave coefficients are well known to be 
non-negative by the no-ghost theorem. A direct proof of unitarity was given in~\cite{Arkani-Hamed:2022gsa,Mansfield:2025gca}~\footnote{The proof is done for $\binom{t+k-1}{k}$, it can be expanded in Legendre polynomial with positive coefficients. But one can easily show that the extra factor in our case $(t+k)$ does not change the positivity of coefficients.}.

$C_{n,k}(t)$ is the correction factor, when $n\to \infty$, $C_{n,k}(t)=1$. Under $t = \frac{k}{2}(\cos\theta-1)$, each factor in $C_{n,k}(t)$ becomes
\begin{equation}
  \frac{n-l}{n-l-t} \;=\; \frac{z_l - 1}{z_l - \cos\theta}\,,
  \qquad z_l \equiv 1 + \frac{2(n-l)}{k}\,.
\end{equation}
Since $n-l\ge 1$, $z_l > 1$ for every $l$. We have Legendre expansion
\begin{equation}
  \frac{z_l-1}{z_l - \cos\theta} 
  = (z_l-1)\sum_{j=0}^{\infty}(2j+1)\,Q_j(z_l)\,P_j(\cos\theta)\,. \label{Q}
\end{equation}
where $Q_j(z_l)$ is a Legendre function of the second kind. For $z_l > 1$, all coefficients $(z_l-1)(2j+1)\,Q_j(z_l)$ are strictly positive.

All factors in $C_{n,k}$ and $V_k$ have a Legendre expansion with positive coefficients, i.e. $C_{n,k}, V_k\in\Lp$. Using the fact that $\Lp$ is closed under products, $R_{n,k} = V_k \cdot C_{n,k}$ also lies in $\Lp$. Hence Eq.~\eqref{eq:amplitude} is a unitary amplitude.


\subsection{IST amplitudes living on the boundary}
For the boundary amplitudes with poles $\{0,1,2,\ldots,n,r\}$ where $r\geq n+1$, positivity at the first $n$ poles can be proved easily. The residue at $s=k$ is just Eq.~\eqref{PF} with $n\to n+1$ times a factor 
\begin{equation}
    \frac{r(r-s-t)}{(r-s)(r-t)}\Bigg/\frac{(n+1)(n+1-s-t)}{(n+1-s)(n+1-t)}\Bigg|_{s=k}.
\end{equation}
We can decompose this factor into partial fractions and use Eq.~\eqref{Q}, its coefficients after the Legendre expansion are
\begin{equation}
    c_{r,j} \;=\; c_+\!\left[\,\delta_{j0} \;+\; \lambda\,(2j+1)\,
\bigl(Q_j(m_1)-Q_j(m_2)\bigr)\right] \;>\; 0\,,
\qquad j=0,1,2,\dots
\end{equation}
with $c_+ \;=\; \frac{r\,(n+1-k)}{(n+1)\,(r-k)} \;>\; 0$,  $\lambda \;=\; \frac{2\,(r-n-1)}{\,r-n-1+k\,} \;>\; 0$ and $m_1 \;=\; 1+\frac{2(n+1-k)}{k}\,, m_2 \;=\; 1+\frac{2r}{k}\,$. $c_{r,j}>0$ because $Q_j(m)$ is a decreasing function for $m>1$ and $m_2>m_1>1$. Since the remaining part~\eqref{PF} has already been shown to lie in $\Lp$, the product of them also lie in $\Lp$.

For the pole $s=r$, the proof is more subtle,
\[
  {\rm Res}_{s=r}A=\frac{r+t}{r-t}\prod_{k=1}^{n}\frac{k(k-r-t)}{(k-r)(k-t)} .
\]
We use throughout
\begin{equation}
  y:=1+\frac tr=\frac{1+\cos\theta}{2}\in[0,1],
\end{equation}
so that partial waves are expanded in $P_j(\cos\theta)=P_j(2y-1)$. Then
\begin{equation}
  {\rm Res}_{s=r}A=\Bigl(\prod_{k=1}^{n}\frac{k}{r-k}\Bigr)\,N(y)\,K(y),
  \label{eq:res}
\end{equation}
with a manifestly positive prefactor and
\begin{equation}
  N(y):=\prod_{k=0}^{n}\Bigl(y-\frac kr\Bigr),\qquad
  K(y):=\frac{1}{2-y}\prod_{k=1}^{n}\frac{1}{1+\frac kr-y} .
  \label{eq:NK}
\end{equation}
We will use four facts: $\Lp$ is closed under sums; $\Lp$ is closed under products; $\Lp$ is closed under $d/dy$, since
$\frac{d}{dy}P_j(2y-1)=\sum_{l<j,\ j-l\ \mathrm{odd}}(2l+1)P_l(2y-1)$; and
$y=\frac12\bigl(1+P_1(2y-1)\bigr)\in\Lp$, hence $y^q\in\Lp$.

\noindent\textbf{\boldmath The numerator $N(y)$:}

Set $a:=\frac{r}{n+1}\ge1$ and $G(y):=\prod_{k=0}^{n}\bigl((n+1)y-k\bigr)$.
\begin{equation}
  N(y)=r^{-n-1}\prod_{k=0}^{n}(ry-k)=r^{-n-1}\,G(ay),
  \label{eq:dil}
\end{equation}
where $ry=(n+1)(a y)$ was used. First, $G(y)\in\Lp$: it is nothing but Veneziano residue in disguise~\footnote{Up to a positive constant.}, which has been proved to have positive Legendre expansion coefficients~\cite{Arkani-Hamed:2022gsa}. 
 
Second, $\Lp$ is stable under $y\mapsto ay$ for $a\ge1$:
\[
  G(ay)=G\bigl(y+(a-1)y\bigr)=\sum_{q}\frac{(a-1)^q}{q!}\,y^q\,G^{(q)}(y),
\]
a sum in which each $G^{(q)}\in\Lp$, each $y^q\in\Lp$, each product $y^qG^{(q)}\in\Lp$,
and each weight $(a-1)^q/q!$ is non-negative because $a\ge1$. With \eqref{eq:dil},
$N(y)\in\Lp$.

\noindent\textbf{\boldmath The denominator $K(y)$:}

$K$ is a finite product of factors $1/(p-y)$ whose poles $p=2$ and $p=1+\frac kr$ all
exceed $1$. For $p>1$,
\[
  \frac{1}{p-y}=\frac{2}{(2p-1)-(2y-1)}=2\sum_{j\ge0}(2j+1)\,Q_j(2p-1)\,P_j(2y-1).
\]
Since $Q_j(2p-1)>0$ for $2p-1>1$. $1/(p-y)$ in $\Lp$, their products
lie in $\Lp$.

Therefore, for general $n$ and $r\geq n+1$, the extremal amplitudes Eq.~\eqref{bamp} are unitary.

\subsection{Critical dimensions from numeric check}
In higher dimensions, the requirement for unitarity is usually stricter. Below we show the critical dimension for IST amplitudes with equidistant poles Eq.~\eqref{eq:amplitude}.

Eq.~\eqref{eq:amplitude} has residue at $s=k$
\begin{equation}
  R_{n,k}(t) = \binom{t+k}{k}
  \;\cdot\;\prod_{l=0}^{k-1}\frac{n-l}{n-l-t}. 
\end{equation}
Set $t=\tfrac{k}{2}(\cos\theta-1)$, and define the
partial-wave coefficients at $s=k$ by the Gegenbauer projection
\begin{equation}
c_{j}^{(n,k)}(d)\;\propto\;\int_{0}^{\pi}\!d\theta\,\sin^{d-3}\theta\,
C^{(\frac{d-3}{2})}_{j}(\cos\theta)\,R_{n,k}\big(t(\cos\theta)\big)\,,
\label{eq:pw}
\end{equation}
with a positive normalization.
 
Scanning all coefficients numerically, we found no critical dimension for $n=1,2$. We give an analytic proof in Appendix~\ref{appC}. We find that for every $n\geq k\ge 3$ the \emph{first} coefficient
to turn negative as $d$ increases is the spin-zero partial wave at $k=3$, $c_{0}^{(n,3)}$, we determine the critical
dimension by the single condition
\begin{equation}
c_{0}^{(n,3)}\big(d_c\big)=0\,.
\label{eq:crit}
\end{equation}
Use the known integral $\int_{-1}^{1} (1-x^2)^{\frac{d-4}{2}} \frac{dx}{1-mx}
= \frac{\sqrt{\pi}\,\Gamma\!\left(\frac{d}{2}-1\right)}{\Gamma\!\left(\frac{d-1}{2}\right)}\,
{}_2F_1\!\left(1,\tfrac{1}{2};\tfrac{d-1}{2};m^2\right)$, we can derive
\begin{equation}
\boxed{\;
c_{0}^{(n,3)}(d)\;=\;\frac{\sqrt{\pi}\;\Gamma\!\big(\tfrac{d}{2}-1\big)}{\Gamma\!\big(\tfrac{d-1}{2}\big)}
\left[\,-1+\sum_{i=0}^{2}\frac{2\,r_{i}}{2(n-i)+3}\;
{}_{2}F_{1}\!\Big(1,\tfrac12;\tfrac{d-1}{2};\Big(\tfrac{3}{2(n-i)+3}\Big)^{\!2}\Big)\right],
\;}
\label{eq:a30closed}
\end{equation}
where $r_{i}=\frac{(-1)^{i}}{2}\binom{2}{i}\,(n-i+1)_{3}$,  with $(\cdot)_3$ the Pochhammer symbol. The critical dimension is listed in Tab.~\ref{tab:dc}
 
\textbf{Large-$n$ limit}
As $n\to\infty$ the level-three residue reduces to the Veneziano value
$\tfrac16(t+1)(t+2)(t+3)=\tfrac{1}{16}(9x^{2}-1)(x+1)$, giving
\begin{equation}
c_{0}^{(\infty,3)}(d)\;\propto\;\;\frac{10-d}{d-1}\,,
\label{eq:veneziano}
\end{equation}
This can also be seen from large $n$ expansion of $a_0^{(n,3)}(d)$. Hence for $n\to\infty$, $d_{c}=10$ as expected since at large $n$, the amplitude in Eq.~\eqref{eq:amplitude} is just Veneziano amplitude.
\begin{table}[htp]
\centering
\small
\setlength{\tabcolsep}{5pt}
\begin{tabular}{l*{8}{c}}
\toprule
$n$        & 3 & 4 & 5 & 6 & 7 & 8 & 9 & 10 \\
$d_{c}$ & 13.159834 & 12.168560 & 11.649282 & 11.328778
           & 11.111356 & 10.954331 & 10.835716 & 10.743025 \\
\midrule
$n$        & 11 & 12 & 13 & 14 & 15 & 16 & 32 & $\infty$ \\
$d_{c}$ & 10.668643 & 10.607661 & 10.556777 & 10.513688
           & 10.476739 & 10.444712 & 10.213742 & 10\\
\bottomrule
\end{tabular}
\caption{Critical spacetime dimension $d_{c}$ of the amplitude \eqref{eq:amplitude}, defined as the
largest $d$ for which all partial-wave coefficients \eqref{eq:pw} are non-negative. In every case,
$d_{c}$ is set by the condition that level-three, spin-zero coefficient $c_{0}^{(n,3)}(d_c)=0$,
Eq.~\eqref{eq:a30closed}. For
$n=1,2$, unitarity holds in all $d$. The values decrease with $n$ and
approach $d_{c}=10$.}
\label{tab:dc}
\end{table}

\section{Full crossing symmetric analogue}\label{sec:closedist}
Inspired by the connection between IST amplitudes and Veneziano amplitude in Eq.~\eqref{FP}, we can also write an IST analogueue in the fully crossing-symmetric case.
\begin{equation}
\boxed{\;A_{\VS,n}(s,t,u)\;=\;-\frac{1}{stu}\prod_{N=1}^{n}\frac{(\mu_N+s)(\mu_N+t)(\mu_N+u)}{(\mu_N-s)(\mu_N-t)(\mu_N-u)}\;,\qquad u=-s-t.\;}
\label{eq:VSN}
\end{equation}
For $\mu_N=N$and $n\to\infty$, this recovers the full Virasoro--Shapiro amplitude. 

There is no {\it a priori} condition like the hidden-zero and splitting conditions that requires such a form. It is simply an analogue of the $s,t$ crossing symmetric case in this paper. This set of IST amplitudes was also recently found in~\cite{Berman:2026ezk} using nonlinear constraints from higher-point factorization and peculiar parity conditions. It would be very interesting to know the connection to the hidden-zero conditions.

The unitarity for equidistant poles is proved analytically in Appendix~\ref{appD}. This is a subset of the triple product form amplitudes~\cite{Huang:2022mdb}. It can be seen as a deformation of the Virasoro–Shapiro amplitude. The Virasoro–Shapiro amplitude was shown to be rigid against deformation~\cite{Geiser:2022exp,Cheung:2022mkw}. This deformation is possible because we gave up on polynomial residues.

For general $\mu_N$, the necessary condition for unitarity used above still holds: $\mu_{N+1}-\mu_{N}\geq\mu_1$ for all $N$. But for pole distributions $\{0,1,2,\ldots,n,r\}$ with 
$r>n+1$, unlike in the $s,t$ crossing symmetric case where $r$ can be as large as one wants, now unitarity requires $r$ not to be too large. One can easily check that the pole distribution $\mu_N\in\{0,1,2,r\}$ breaks unitarity when $r\geq14$. This brings a problem: if we have only poles at $\mu_N=0,1,2$, we cannot add a pole at infinity, which should be irrelevant in the IR by intuition. It seems to suggest that the poles cannot be too dense or too sparse in this case. If we have a pole at infinity, there have to be infinitely many poles distributed such that the density of the poles does not have a big bump. Put another way, the distribution of the poles in the IR imposes strong unitarity constraints on the tree-level UV completion. Indeed, as shown by a dispersion-relation analysis~\cite{Haring:2024wyz}, in the small-impact-parameter region, high spin states as well as high energy levels are needed to reproduce the graviton pole in the IR~\footnote{We would like to thank Anna Tokareva for discussions on this.}.


\subsection{Low energy expansion}\label{subsec:VSWilson}
We want to know where these IST amplitudes lie in the fully crossing-symmetric massless scalar positivity bounds in~\cite{Caron-Huot:2020cmc}. Take the low energy ansatz for 4-point amplitudes
\begin{align}
A_{\mathrm{low}}(s,t)
={}-&
\frac{g^2}{s\, t \,u}
-\lambda \nonumber\\
&+g_2\left(s^2+t^2+u^2\right)
+g_3(stu)
+g_4\left(s^2+t^2+u^2\right)^2
+g_5\left(s^2+t^2+u^2\right)(stu)+\cdots.
\end{align}
We will set $g=1$ for convenience. For the IST amplitudes with poles at $\{0,1,2,\ldots,n\}$
\begin{equation}
\boxed{\;g_2^{(n)}=-H_n^{(5)},\quad g_3^{(n)}=-2(H_n^{(3)})^2,\quad g_4^{(n)}=-\tfrac12 H_n^{(7)}.\;}
\label{eq:Vstildecoeff}
\end{equation}

In order to know the location of Eq.~\eqref{eq:VSN} with $\mu_N\in \{0,1,2,\ldots,n\}$, we need to subtract the spin-0 part of the amplitude as in~\cite{Caron-Huot:2020cmc}. This is shown in Appendix~\ref{appE}. The scalar-subtracted Wilson coefficients are
\begin{equation}
\begin{aligned}
g_2^{\rm ss}(n)&=-H_n^{(5)}+2\,\mathcal T_3^{(n)},\\
g_3^{\rm ss}(n)&=-2(H_n^{(3)})^2+6\,\mathcal T_4^{(n)},\\
g_4^{\rm ss}(n)&=\tfrac12\bigl[-H_n^{(7)}+2\,\mathcal T_5^{(n)}\bigr],
\end{aligned}
\end{equation}
where
\begin{equation}
\mathcal T_p^{(n)}\;=\!\!\!\!\sum_{\substack{1\le l,k\le n\\ l+k\ge n+1}}\!\!\!\!\frac{A_l^{(n,k)}}{k^{p+1}}\ln\!\Bigl(1+\frac{k}{l}\Bigr),
\end{equation}
with
\begin{equation}
A_l^{(n,k)}\;=\;\frac{(-1)^{k+n-l+1}(n+k)!(n+l)!\,[(k+l-1)!]^2}{(k!)^2(l!)^2(n-k)!(n-l)!(k+l-n-1)!(n+k+l)!}.
\label{eq:Al}
\end{equation}

\begin{figure}
    \centering
    \includegraphics[width=0.7\linewidth]{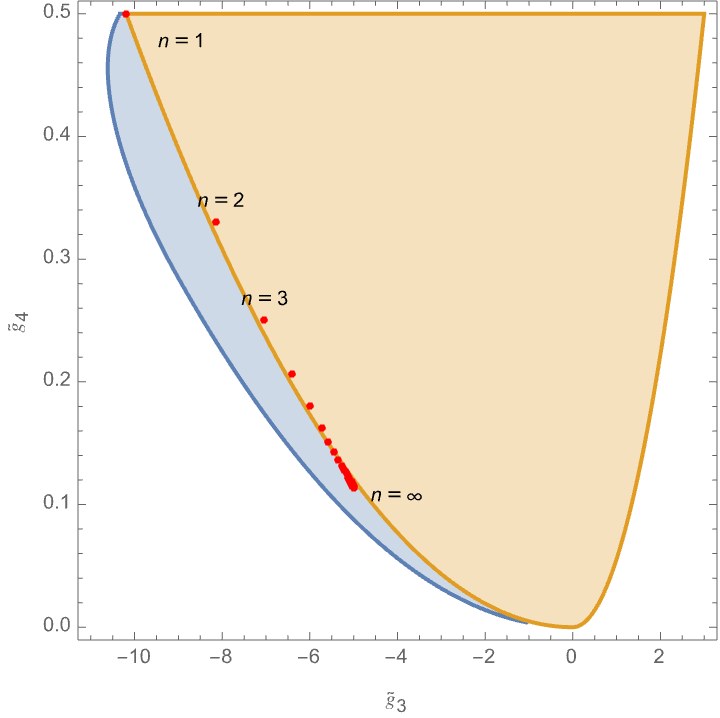}
    \caption{Location of amplitudes Eq.~\eqref{eq:VSN} with $\mu\in\{0,1,2,\ldots,n\}$. $n=1$ is on the kink, $n=\infty$ is the scalar-subtracted Virasora--Shapiro amplitude} 
    \label{VS}
\end{figure}

The points $\bigl(\gt_3(n),\gt_4(n)\bigr)$ are shown in Fig.~\ref{VS}, where $\gt_3(n)=g_3^{\rm ss}(n)/g_2^{\rm ss}(n)$, $\gt_4(n)=g_4^{\rm ss}(n)/g_2^{\rm ss}(n)$.

\begin{center}
\begin{tabular}{c|c|c|c|c}
$n$ & $\gt_3$ & $\gt_4$ & region\\\hline
1 & $-10.19196$ & $0.50000$ & kink\\
2 & $-8.14096$  & $0.32949$ & yellow\\
3 & $-7.05524$  & $0.25085$ & yellow\\
5 & $-6.00916$  & $0.17993$ & yellow\\
8 & $-5.43796$  & $0.14256$ & yellow\\
9 & $-5.34490$  & $0.13654$ & {\bf blue}\\
20 & $-5.02922$ & $0.11625$ & {\bf blue}\\
$\infty$ & $-4.93630$ & $0.11030$ & {\bf blue}
\end{tabular}
\end{center}
The left boundary of the yellow region is a parabola. It is the mass scaling curve of scalar-subtracted $n=1$ IST amplitude with spin-0 partial wave subtracted
\begin{equation}
    A(s,t)=-\frac{1}{s t u}\frac{(m+s)(m+t)(m+u)}{(m-s)(m-t)(m-u)}+c^{(1,1)}_0M^{(m)}_{spin-0}\quad \text{for}\quad m\in[1,\infty], 
\end{equation}
where $M^{(m)}_{spin-0}=1/(m-s)+1/(m-t)+1/(m-u)$, and $c^{(1,1)}_0$ is the spin-zero coefficient of the first term. The trajectory of IST amplitudes with pole distribution $\{0,1,2,\ldots,n\}$ crosses the parabola
\emph{between $n=8$ and $n=9$}. As can be seen in Fig.~\ref{VS}.
The endpoint $n=1$ sits on the kink point. The infinite limit lands inside the blue sliver at the scalar-subtracted Virasoro–Shapiro point. It is known that scalar-subtracted Virasoro–Shapiro amplitudes, DBI amplitudes and their deformations lie inside the blue sliver~\cite{Berman:2025owb}. However, it is still an open question to find a closed-form amplitude that lives on the boundaries. What has been understood is that, using SDPB, maximizing the lowest-lying spin-2 coupling, one can reach the extremal boundaries. The spectrum on the boundary shows a nonlinear leading trajectory with spurious high-spin states~\cite{Haring:2023zwu,Albert:2023seb,Berman:2024eid,Albert:2024yap}. The spurious poles are really needed. (A single trajectory was shown to be incompatible with crossing symmetry and polynomial boundedness~\cite{Eckner:2024ggx,Eckner:2024pqt,Eckner:2025kve})

\begin{figure}
    \centering
    \includegraphics[width=0.7\linewidth]{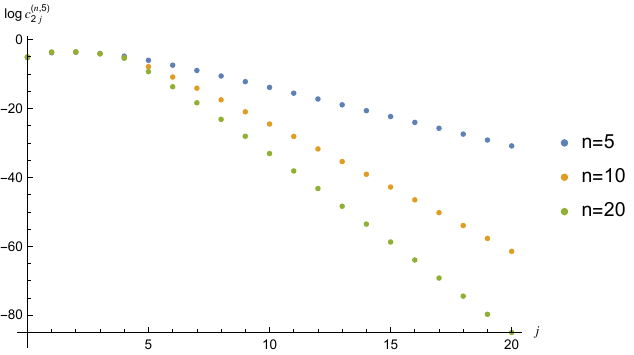}
    \caption{Logarithm of partial wave coefficients for different $n$ at $s=5$.}
    \label{LSD}
\end{figure}

The IST amplitudes do exhibit low-spin dominance, like many amplitudes. This can be seen in Fig.~\ref{LSD}. We fix $s=5$; for different $n$, all higher spins are exponentially suppressed, and larger $n$ gives a larger slope. In the $n\to \infty$ limit, the amplitude becomes the Virasoro–Shapiro amplitude, and coefficients with $j>8$ vanish exactly. Furthermore, one of the best-known IST amplitudes --- $n=1$ --- sits on the kink point that connects to the extremal boundary. So it is possible that the extremal amplitudes lying on the boundary can be found by deforming IST amplitudes, and this provides the needed high-spin states with exponentially suppressed couplings~\cite{Chiang:2022jep}. We will leave this for future work.

If we insist only on low-spin dominance and do not require polynomial residues, there could be many interesting amplitudes constructed from different distributions of the poles. For example, in the $s,t$ crossing symmetric case, if we have such a pole distribution $\mu_N=N(N+1/2)$ for $n\in\{1,2,3,\ldots,\infty\}$, the amplitude in Eq.~\eqref{GIST} will give a closed-form expression
\begin{equation}
    -\frac{2\pi\sin\pi\lambda(s+t)}{\sin\pi\lambda(s)\,\sin\pi\lambda(t)},
\end{equation}
where $\lambda(s)=\frac{1}{2}(\sqrt{1+8s}-1)$. The nonlinear $\lambda(s)$ is reminiscent of the "bespoken" amplitudes~\cite{Cheung:2023uwn}. In the $s,t,u$ symmetric case, the same spectrum gives
\begin{equation}
    \frac{1}{stu}
\frac{\sin\pi\lambda(-s)}{\sin\pi\lambda(s)}\frac{\sin\pi\lambda(-t)}{\sin\pi\lambda(t)}\frac{\sin\pi\lambda(-u)}{\sin\pi\lambda(u)}.
\end{equation}
This is an example that meromorphic amplitudes with nonlinear spectrum that generates graviton pole in the IR.

\section{Conclusion}\label{sec:conclusion}

In this paper, we studied the properties of IST amplitudes. In the
$s,t$ crossing-symmetric case, the hidden-zero and splitting conditions
impose highly non-trivial constraints on four-point amplitudes. They
restrict the amplitudes to the form
$
    \frac{f(s)f(t)}{f(s+t)}
$,
a class that is closed under multiplication. In the product
representation, owing to crossing symmetry, polynomial boundedness, and
causality, the simplest factor one can write is
\[
    \frac{\mu_n-s-t}{(\mu_n-s)(\mu_n-t)}.
\]
We found the maximal allowed ``rule-in'' region arising from this
meromorphic amplitude construction. It corresponds to the pole distribution
$    \{0,1,2,\ldots,n,r\}$, with $r\geq n+1$.
The unitarity of these amplitudes are proved analytically. There is still a gap
between the ``rule-in'' and ``rule-out'' regions. Given the reasonable
assumptions made so far, we think that this could have two possible
explanations: one is that the energy levels have accumulation points, as in
the Coon amplitude; the other is that numerical rule-out bounds has not achieved convergence with respect to nonlinear constraints Eq.~\eqref{fac}. It would be instructive to extract the extremal spectrum associated with the solution on the dual bootstrap boundary~\cite{Berman:2025owb} and compare it with the analytic spectra constructed in this work.

It is straightforward to generalize the construction to the fully
$s,t,u$ crossing-symmetric case. These $s,t,u$ crossing-symmetric IST
amplitudes have a triple-product form~\cite{Huang:2022mdb}. They were also
recently found in Ref.~\cite{Berman:2026ezk} using nonlinear constraints
from higher-point factorization and a peculiar parity condition. It would
be very interesting to understand their connection to the hidden-zero
conditions. A triple-product ansatz, combined with bootstrap constraints,
also uniquely determines the Virasoro--Shapiro amplitude if one assumes
infinitely many energy levels~\cite{Xu:2026kix}.

The unitarity constraints on the distribution of poles are more intricate
in the fully crossing-symmetric case. They require the poles to be neither
too close to nor too far from one another. Adding a pole at infinite energy
would require the presence of infinitely many poles at intermediate
energies. This is compatible with the stringy description of the
generation of the graviton pole in the IR~\cite{Haring:2024wyz}. If we
require only low-spin dominance and do not impose polynomial residues,
there could be a much broader class of amplitudes. With recent developments
in the bootstrap of gravity
\cite{Arkani-Hamed:2021ajd,Sinha:2020win,Caron-Huot:2021rmr,Haring:2022cyf,Cheung:2024obl,Berman:2024owc,Chang:2025cxc,Tokareva:2025rta,Bellazzini:2025bay,Peng:2026ztp,Berman:2026ezk},
it would be interesting to determine where these IST amplitudes lie and how
meromorphic IST amplitudes fare under full unitarity
\cite{Paulos:2017fhb,Correia:2020xtr,Guerrieri:2020bto,
Chen:2022nym,deRham:2026lvc}.

\begingroup
\renewcommand{\addcontentsline}[3]{}
\begin{acknowledgments}
We thank Justin Berman, Mariana Carrillo González, Aidan Herderschee, Chang Hu, Alexey Koshelev, Guangzhuo Peng, Anna Tokareva, Yongjun Xu, Roman Zwicky for valuable discussions and comments on the draft. LQS is grateful to Roman Zwicky for the kind hospitality during the visit to the University of Edinburgh. LQS and AV are partially supported by the Italian MUR under the project 20223ANFHR (PRIN2022). Part of this work was completed during the workshop “Primordial Cosmology” at the Munich Institute for Astro-, Particle and BioPhysics (MIAPbP).
\end{acknowledgments}
\endgroup

\appendix



\section{General solution of the cocycle equation}
\label{appA}

In this appendix, we show that the general meromorphic solution of the
cocycle equation
\begin{equation}\label{eq:cocycle}
  A(a,\, b+c)\,A(b,\,c)\;=\;A(a,\,b)\,A(a+b,\,c),
\end{equation}
is
\begin{equation}\label{eq:cocycle-sol}
  A(a,b)\;=\;C\,\frac{f(a)\,f(b)}{f(a+b)}\,,
\end{equation}
with $f$ meromorphic on $\mathbb{C}$. We assume
throughout that $A$ is meromorphic on $\mathbb{C}^2$ and
$\mathcal{A}\not\equiv0$, and that \eqref{eq:cocycle} holds for all complex number
$a,b,c$.

\medskip
\noindent\textit{Proof.}
It is convenient to work not with $\log\mathcal{A}$, which is multivalued,
but with the logarithmic derivative
\begin{equation}\label{eq:Ldef}
  L(a,\,b)\;\equiv\;\frac{\partial_2\,A(a,\,b)}{A(a,\,b)}\,,
\end{equation}
which is an ordinary meromorphic function of the two variables.

Taking the logarithmic derivative with respect to $c$ of \eqref{eq:cocycle} gives
\begin{equation}\label{eq:L-cocycle}
  L(a,\,b+c)+L(b,\,c)\;=\;L(a+b,\,c)\,.
\end{equation}
Now set $c=\varepsilon$, where $\varepsilon$ is any fixed value such that
$A(\cdot,\varepsilon)$ is neither identically zero nor
identically infinite~\footnote{One cannot take
$\varepsilon=0$: if $A$ has a pole at $t=0$ then
$L(x,0)\equiv\infty$.}.
Writing $h(x)\equiv L(x-\varepsilon,\,\varepsilon)$ and shifting
$b\to b-\varepsilon$, eq.~\eqref{eq:L-cocycle} becomes
\begin{equation}\label{eq:L-solved}
  L(a,\,b)\;=\;h(a+b)\,-\,h(b)\,.
\end{equation}

Let $f$ be the solution of
\begin{equation}\label{eq:fdef}
  \frac{f'}{f}\;=\;-h\,,\qquad\text{i.e.}\qquad
  f(x)\;=\;\exp\Big[-\!\int^{x}\! h(y)\,dy\Big]\,,
\end{equation}
which is fixed up to a multiplicative constant. Because $A(a,b)$ is meromorphic, its logarithmic derivative $L(a,b)$ is also meromorphic with integer residues on both variables. Fix $b$ to be some constant that makes $h(b)$ regular, Eq.~\eqref{eq:L-solved} means $h$ function is also meromorphic with integer residues. From Eq.~\eqref{eq:fdef}, we know that $f$ is also a meromorphic function. It has poles and zeros only at the poles of $h$.

Using $h=-f'/f$, eq.~\eqref{eq:L-solved} states
that $A(a,b)$ and $f(b)/f(a+b)$ have the same logarithmic
derivative in $b$. Their ratio is therefore independent of $b$,
\begin{equation}\label{eq:A-R}
  A(a,\,b)\;=\;R(a)\,\frac{f(b)}{f(a+b)}\,.
\end{equation}

Finally we fix $R$ by substituting \eqref{eq:A-R} back
into \eqref{eq:cocycle}:
\begin{equation}
\frac{R(a+b)}{f(a+b)}=\frac{R(b)}{f(b)}.
\end{equation}
The right-hand side does not depend on $a$, i.e. is independent of its argument. So $R(a)/f(a)$ is
a constant $C$, and \eqref{eq:A-R} becomes \eqref{eq:cocycle-sol}.

\medskip
\noindent\textit{Remark }
For a given $A$ the function $f$ is unique only up to
$f(x)\to\lambda\,e^{\alpha x}f(x)$, which leaves \eqref{eq:cocycle-sol}
invariant. More importantly, the cocycle equation is far from fixing the
amplitude: for \emph{any} entire function $q$,
\begin{equation}\label{eq:exp-freedom}
  f\;\to\;f\,e^{\,q}
  \qquad\Longrightarrow\qquad
  \mathcal{A}(s,t)\;\to\;\mathcal{A}(s,t)\,e^{\,q(s)+q(t)-q(s+t)}.
\end{equation}
This maps solutions to solutions. The choice $q(x)=-\tfrac{\beta}{2}x^2$
reproduces the factor $e^{\beta st}$, which indeed satisfies
\eqref{eq:cocycle} on its own. Therefore to select a unique amplitude $A(s,t)$, one needs to implement polynomial boundedness on the large-$|s|$ growth of the amplitude.

\section{Boundary pole distribution}\label{appB}
We will prove the following claim in this appendix. Fix $Y=\sum_N \mu_N^{-2}\in [H_n^{(2)}, H_{n+1}^{(2)}]$,
the spectrum satisfies $\mu_1 = 1$ and $\mu_{N+1}-\mu_N \geq 1$ for all
$N \geq 1$, then
\begin{equation}
  X := \frac{\sum_{N} \mu_N^{-3}}{Y}
  \;\leq\;
  \frac{H_n^{(3)} + \bigl(Y - H_n^{(2)}\bigr)^{3/2}}{Y}.
  \label{eq:Xbound}
\end{equation}
And the boundary is reached by the spectrum with smallest number of poles
\begin{equation}
  \mu_N =
  \begin{cases}
    N, & 1 \leq N \leq n, \\[4pt]
    \bigl(Y - H_n^{(2)}\bigr)^{-1/2}, & N = n+1, \\[4pt]
    \infty, & N > n+1.
  \end{cases}  \label{pd}
\end{equation}
It is easy to see that this pole distribution saturates the boundary $X\,Y=\sum_N \mu_N^{-3} = H_n^{(3)} + (Y - H_n^{(2)})^{3/2}$.

\medskip
For any other ordered distribution $\{0,\nu_1,\nu_2,\ldots,\nu_m\}$, we will show that 
\begin{equation}
    \sum_k(\nu_k^{-3}-\mu_k^{-3})\leq0,
\end{equation}
so that the pole distribution Eq.~\eqref{pd} gives larger $X$ than $\{0,\nu_1,\nu_2,\ldots,\nu_m\}$. 

Use the following inequality
\begin{equation}
  \nu^{-3} - \mu^{-3}
  \;\leq\;
  \tfrac{3}{2}\, \nu^{-1} \bigl(\nu^{-2} - \mu^{-2}\bigr).
  \label{eq:elem}
\end{equation}
Then we have (This comes from the convexity of the function $y(x) = x^{3/2}$: applying
$y(x) \geq y(x_0) + y'(x_0)(x - x_0)$ at $x = \mu^{-2}$, $x_0 = \nu^{-2}$)
\begin{equation}
    \sum_k(\nu_k^{-3}-\mu_k^{-3})\leq\sum_k \tfrac{3}{2}\, \nu_k^{-1} \bigl(\nu_k^{-2} - \mu_k^{-2}\bigr). \label{B5}
\end{equation}

Define
\begin{equation}
  P_K := \sum_{k \leq K} \bigl(\nu_k^{-2} - \mu_k^{-2}\bigr),
\end{equation}
From the necessary unitarity condition $\nu_k-\nu_{k-1}\geq1$, it is easy to deduce that $\nu_k\geq k$, so we have $P_K \leq 0$ for all $K$, and $P_K \to 0$ as $K \to \infty$
since both spectra sum to $Y$. Apply $\nu_k^{-2} - \mu_k^{-2} = P_k - P_{k-1}$ on right hand side of Eq.~\eqref{B5},
\begin{equation}
  \sum_{k \leq K} \nu_k^{-1} \bigl(\nu_k^{-2} - \mu_k^{-2}\bigr)
  = \sum_{k \leq K-1} \bigl(\nu_k^{-1} - \nu_{k+1}^{-1}\bigr) P_k
    + \nu_K^{-1} P_K.
\end{equation}
Take $K \to \infty$, each term of the sum is nonpositive, since $\nu_k^{-1} - \nu_{k+1}^{-1} > 0$ and $P_k \leq 0$; the second term on the right hand side vanishes, so
\begin{equation}
  \sum_{k} \bigl(\nu_k^{-3} - \mu_k^{-3}\bigr)
  \;\leq\;
  \tfrac{3}{2} \sum_{k} \nu_k^{-1} \bigl(\nu_k^{-2} - \mu_k^{-2}\bigr)
  \;\leq\; 0.
\end{equation}
Therefore the $X$ for pole distribution Eq.~\eqref{pd} is the largest boundary that saturates the inequality Eq.~\eqref{eq:Xbound}.

\section{Unitarity for $n=1,2$, $s,t$ crossing symmetric IST amplitudes in any dimension}\label{appC}


Consider the amplitude
\begin{equation}\label{eq:An}
  A_n(s,t)=-\frac{s+t}{s\,t}\prod_{N=1}^{n}\frac{N\,(N-s-t)}{(N-s)(N-t)}\,.
\end{equation}
The residues $R_{n,k}$ for fixed $n$ and $s=k$ are
\begin{equation}\label{eq:res12}
R_{1,1}=\frac{1+\cos\theta}{3-\cos\theta}\,,
  \quad
R_{2,1}=\frac{2(1+\cos\theta)}{5-\cos\theta}\,,\quad
  R_{2,2}=\frac{1+\cos\theta}{3-\cos\theta}\cdot\frac{\cos\theta}{2-\cos\theta}\,.
\end{equation}

The partial waves expansion in $d$ dimensions is defined by
$R_{n,k}=\sum_j c^{(n,k)}_j\,C_j^{(\lambda)}(\cos\theta)$ with
$\lambda=\frac{d-3}{2}$, so that
\begin{equation}\label{eq:aj}
  a_j^{(n,k)}\;\propto\;\int_0^{\pi}\!d\theta\,\sin^{d-3}\!\theta\;
  C_j^{(\lambda)}(\cos\theta)\,R_{n,k}\,,
\end{equation}
with a positive proportionality constant. We show that $c^{(n,k)}_j>0$ for every
$j$ and every $d$, so for $n=1,2$, Eq.~\eqref{eq:An} has no critical dimension.

All poles of \eqref{eq:aj} sit at $\cos\theta=x_0\in\{2,3,5\}$, i.e.\ outside the physical region $|\cos\theta|\le1$.
Reducing each residue,
\begin{equation}\label{eq:pf}
  \frac{1+\cos\theta}{x_0-\cos\theta}=-1+\frac{1+x_0}{x_0-\cos\theta}\,,
  \qquad
  \frac{\cos\theta\,(1+\cos\theta)}{(2-\cos\theta)(3-\cos\theta)}
  =1+\frac{6}{2-\cos\theta}-\frac{12}{3-\cos\theta}\,.
\end{equation}

Since $x_0\ge2$, the Taylor expansion
\begin{equation}\label{eq:geo}
  \frac{1}{x_0-\cos\theta}=\sum_{m\ge0}\frac{\cos^m\theta}{x_0^{m+1}},
\end{equation}
converges on $\theta\in[0,\pi]$. Inserting \eqref{eq:geo} into \eqref{eq:pf}
gives expansions with manifestly positive Taylor coefficients:
\begin{align}
  \frac{1+\cos\theta}{x_0-\cos\theta}
  &=\frac{1}{x_0}+\sum_{m\ge1}\frac{1+x_0}{x_0^{m+1}}\cos^m\theta,
  \qquad x_0=3,5\,,\label{eq:t1}\\[4pt]
  \frac{\cos\theta\,(1+\cos\theta)}{(2-\cos\theta)(3-\cos\theta)}
  &=\sum_{m\ge1}\Big(\frac{3}{2^m}-\frac{4}{3^m}\Big)\cos^m\theta\,,
  \qquad \frac{3}{2^m}-\frac{4}{3^m}>0
  \ \ \text{since }\Big(\tfrac32\Big)^m\ge\tfrac32>\tfrac43\,.\label{eq:t2}
\end{align}

Then use the fact that $\cos^m\theta$ can be expanded in terms of Gegenbauer polynomial with positive coefficients in any dimension, so does the positive linear combination of different powers of $\cos\theta$. Therefore the residues in Eq.~\eqref{eq:t1} and~(\ref{eq:t2}) can be expanded in Gegenbauer polynomial with positive coefficients in any dimension and Eq.~\eqref{eq:An} has no critical dimensions.

\section{Unitarity for fully crossing-symmetric IST amplitudes}\label{appD}
  The residue at $s=k$ takes the form
\begin{equation}\label{eq:Ffinal}
  \boxed{R_{n,k}(t) = 
  \underbrace{\frac{\bigl[(t+1)(t+2)\cdots(t+k-1)\bigr]^2}{(k!)^2}}_{\displaystyle \mathrm{VS}_k(t)}
  \;\cdot\;
  \underbrace{\prod_{l=0}^{k-1}
  \frac{(n-l)(n+k-l)}{(n-l-t)(n-l+k+t)}}_{\displaystyle D_{n,k}(t)}\,,}
\end{equation}
where $\mathrm{VS}_k(t)$ is the residue of the Virasoro-Shapiro 
amplitude at $s=k$. This holds for $k>1$.
For $k=0$ the residue is $1/t^2$, the graviton pole, and for $k=1$, $VS_1(t)=1$. It is the square of the residues of Veneziano amplitudes, therefore it has positive Legendre expansion coefficients which follows from the unitarity of Veneziano amplitude~\cite{Arkani-Hamed:2022gsa}. The analytic proof here is for $D\leq10$, since Veneziano amplitude is unitary for $D\leq10$.

Under the change of variable to $\cos\theta=1+2t/k$, 
the correction factor $D_{n,k}$ becomes
\begin{equation}\label{eq:Dx}
  D_{n,k}(t) = \prod_{l=0}^{k-1}\frac{z_l^2-1}{z_l^2-\cos^2\theta}\,,
  \qquad z_l = 1+\frac{2(n-l)}{k} > 1\,.
\end{equation}
The full correction 
$D_{n,k}\to 1$ as $n\to\infty$ since $z_l\to\infty$.

As was used in Eq.~\eqref{Q}
\begin{equation}
  \frac{1}{z-x} = \sum_{j=0}^{\infty}(2j+1)\,Q_j(z)\,P_j(x)\,.
\end{equation}
Writing $\frac{1}{z^2-\cos\theta^2}=\frac{1}{2z}\bigl(\frac{1}{z-\cos\theta}+\frac{1}{z+\cos\theta}\bigr)$ and using $P_j(-\cos\theta)=(-1)^j P_j(\cos\theta)$, the odd-$J$ terms cancel and the even-$j$ terms double. Therefore, for $z_l>1$ the correction factor 
\begin{equation}
    \prod_{l=0}^{k-1}\frac{1}{z_l^2-\cos^2\theta}=\frac{1}{z_l}\prod_{l=0}^{k-1}\sum_{\text{even},j}^{\infty}(2j+1)Q_j(z_l)P_j(\cos\theta),
\end{equation}
has Legendre expansion with positive coefficients. Now very factor of Eq.~\eqref{eq:Ffinal} is either are positive constant or has Legendre expansion with positive coefficients. Therefore, their product also has Legendre expansion with positive coefficients and the amplitude is unitary.

\section{Scalar subtraction of IST amplitudes}\label{appE}
The residue of Eq.~\eqref{eq:VSN} at $s=k$ is
\begin{equation}
 R_{n,k}(t) = 
  \frac{\bigl[(t+1)(t+2)\cdots(t+k-1)\bigr]^2}{(k!)^2}
  \;\cdot\;
  \prod_{l=0}^{k-1}
  \frac{(n-l)(n+k-l)}{(n-l-t)(n-l+k+t)}\,.
\label{eq:VSresidue}
\end{equation}
Changing variables to $x=1+2t/k$,
the spin-0 partial wave coefficients are
\begin{equation}
a_0^{(n,k)}\;=\;\frac{2}{k}\sum_{l=n-k+1}^{n}A_l^{(n,k)}\,\ln\!\Bigl(1+\frac{k}{l}\Bigr),
\label{eq:c0closed}
\end{equation}
with
\begin{equation}
A_l^{(n,k)}\;=\;\frac{(-1)^{k+n-l+1}(n+k)!(n+l)!\,[(k+l-1)!]^2}{(k!)^2(l!)^2(n-k)!(n-l)!(k+l-n-1)!(n+k+l)!}.
\label{eq:Al}
\end{equation}

Subtract the $j=0$ partial wave at every energy level,
\begin{equation}
A_{\VS,n}^{\rm ss}\;=\;A_{\VS,n}\;+\;\sum_{k=1}^{n}a_0^{(n,k)}\,\Mss^{(k)},
\end{equation}
where $\Mss^{(k)}=1/(k-s)+1/(k-t)+1/(k-u)$. It has only spin-zero partial wave and is used to cancel the spin-zero partial wave of $A_{\VS,n}$. We have the scalar-subtracted Wilson coefficients
\begin{equation}
\begin{aligned}
g_2^{\rm ss}(n)&=-H_n^{(5)}+2\,\mathcal T_3^{(n)},\\
g_3^{\rm ss}(n)&=-2(H_n^{(3)})^2+6\,\mathcal T_4^{(n)},\\
g_4^{\rm ss}(n)&=\tfrac12\bigl[-H_n^{(7)}+2\,\mathcal T_5^{(n)}\bigr],
\end{aligned}
\end{equation}
where
\begin{equation}
\mathcal T_p^{(n)}\;=\!\!\!\!\sum_{\substack{1\le l,k\le n\\ l+k\ge n+1}}\!\!\!\!\frac{A_l^{(n,k)}}{k^{p+1}}\ln\!\Bigl(1+\frac{k}{l}\Bigr).
\end{equation}

\bibliography{main}

\end{document}